\documentclass[12pt,preprint]{aastex}

\newcommand{\iso}[2]{\hbox{${}^{#1}{\rm #2}$}}
\newcommand{\Msun}{\ensuremath{{M}_{\sun}}}
\newcommand{\Lsun}{\ensuremath{{L_{\sun}}}}

\usepackage{rotating}

\shorttitle{Nucleosynthesis in the Brightest AGB stars}
\shortauthors{Karakas, Garc{\'{\i}}a-Hern{\'a}ndez, \& Lugaro}

\begin{document}


\title{Heavy Element Nucleosynthesis in the Brightest Galactic 
Asymptotic Giant Branch stars}


\author{Amanda I. Karakas}
\affil{Research School of Astronomy \& Astrophysics, Mount Stromlo Observatory,
Weston Creek ACT 2611, Australia}
\email{akarakas@mso.anu.edu.au}

\author{D. A. Garc{\'{\i}}a-Hern{\'a}ndez\altaffilmark{1}}
\affil{Instituto de Astrofisica de Canarias, C/ Via L\'{a}ctea s/n,
38200 La Laguna (Tenerife), Spain}
\email{agarcia@iac.es}

\and

\author{Maria Lugaro}
\affil{Monash Centre for Astrophysics, Monash University,
Clayton VIC 3800, Australia}
\email{maria.lugaro@monash.edu.au}


\altaffiltext{1}{Departamento de Astrofísica, Universidad de La Laguna (ULL),
E-38205 La Laguna, Spain}


\begin{abstract}
We present updated 
calculations of stellar evolutionary sequences and detailed
nucleosynthesis predictions for the brightest asymptotic giant branch
(AGB) stars in the Galaxy with masses between 5$\Msun$ to 9$\Msun$,
with an initial metallicity of $Z =0.02$ ([Fe/H] = 0.14).  
In our previous studies we used the \citet{vw93} 
mass-loss rate, which stays low until the pulsation period reaches
500 days after which point a superwind begins. 
\citet{vw93} noted that for stars over 2.5$\Msun$ the superwind
should be delayed until  $P \approx 750$ days at 5$\Msun$.
We calculate evolutionary sequences where we delay the 
onset of the superwind to pulsation periods of 
$P \approx 700-800$ days in models of $M=5$, 6, and 7$\Msun$. 
Post-processing nucleosynthesis calculations show
that the 6 and 7$\Msun$ models produce the most Rb, with [Rb/Fe]
$\approx 1$~dex, close to the average of most of the Galactic 
Rb-rich stars ([Rb/Fe] $\approx 1.4 \pm 0.8$~dex).
Changing the rate of the \iso{22}Ne $+ \alpha$ reactions 
results in variations of [Rb/Fe] as large as 0.5~dex in models
with a delayed superwind. The largest enrichment in heavy
elements is found for models that adopt the NACRE rate
of the \iso{22}Ne($\alpha$,n)\iso{25}Mg reaction. Using this
rate allows us to best match the composition of most of the
Rb-rich stars. A synthetic evolution algorithm is 
then used to remove the remaining envelope resulting in 
final [Rb/Fe] of $\approx 1.4$~dex although with C/O ratios 
$>1$. We conclude that delaying the superwind may account
for the large Rb overabundances observed in the brightest
metal-rich AGB stars. 
\end{abstract}





\section{Introduction}

\citet{garcia06} and \citet{garcia09} identified several Galactic 
and Magellanic Cloud intermediate-mass asymptotic giant branch (AGB)
stars within a sample of OH/IR stars (i.e., bright O-rich giants
with large infrared excesses).
The large enhancements of the neutron-capture element rubidium (Rb) 
found by these authors, combined with the fact that these stars are 
O-rich, support the prediction that hot bottom  burning (HBB) and 
an efficient third dredge-up (TDU) have occurred in these stars
\citep{garcia07a}.
The Rb is hypothesized to be produced by the $slow$ neutron-capture 
process (the $s$ process) in intermediate-mass AGB stars, that 
in this context have masses over $M \gtrsim 4\Msun$. 
An enrichment in the element Rb over the elements Sr, Y, and Zr 
is a tantalizing piece of evidence for the efficient operation 
of the \iso{22}Ne($\alpha$,n)\iso{25}Mg neutron source in 
intermediate-mass AGB stars \citep{truran77,cosner80}.
This is because the production of Rb is sensitive to the 
high neutron density associated with this neutron source 
\citep{lambert95,abia01,vanraai12}.  
These observational results are particularly important because there is a 
paucity of observational evidence for constraining stellar models
of intermediate-mass AGB stars, and especially the efficiency 
of the TDU and HBB. These two last points
are much debated in the context of the globular cluster abundance 
anomalies
\citep{karakas02,stancliffe04b,herwig04a,ventura05a,karakas06b}.
Intermediate-mass AGB stars play an important role in chemical
evolution of galaxies and stellar systems by producing substantial
nitrogen \citep[e.g.,][]{romano10,kobayashi11a}. These stars may also
prove to be important for the chemical evolution of a select number of
neutron-capture elements such as Rb \citep{travaglio04}.

Briefly, during the thermally-pulsing AGB phase the He-burning 
shell becomes thermally unstable every $\approx 10^{3-4}$ years 
for stars with H-exhausted core masses over 
$M_{\rm c} \gtrsim 0.8\Msun$ \citep[see][for a review]{herwig05}.
The energy provided by the thermal pulse (TP) expands the whole 
star, pushing the H shell out to cooler regions where it is almost 
extinguished, and subsequently allowing the convective 
envelope to move inwards (in mass) to regions previously 
mixed by the TP driven convective zone. This inward movement of 
the convective envelope is known as the third dredge-up (TDU), and 
is responsible for enriching the surface in \iso{12}C and other 
products of He-burning, as well as heavy elements produced by 
the $s$ process \citep[e.g.,][]{busso99}. In intermediate-mass AGB stars with 
initial masses  $\gtrsim 4\Msun$, the base of the convective 
envelope can dip into the top of the H-burning shell, 
causing proton-capture nucleosynthesis to occur there 
\citep[see][for a review]{lattanzio96}. HBB nucleosynthesis 
converts the newly created \iso{12}C into \iso{14}N and can 
prevent the formation of a C-rich atmosphere where C/O $\ge 1$. 

\citet{vanraai12} performed a detailed study on the production of Rb
in stellar models of intermediate-mass AGB stars covering a range in 
compositions from solar metallicity down to the metallicity 
of the Small Magellanic Cloud
($Z=0.004$ or [Fe/H]\footnote{We adopt the standard
spectroscopic notation [X/Y] = $\log_{10}$(X/Y)$_{\rm star} -
\log_{10}$(X/Y)$_{\odot}$ where X and Y are abundances by number.}
= $-0.7$).  While the qualitative features of the 
observations could be reproduced
(increasing [Rb/Fe] ratio with increasing stellar mass or decreasing
metallicity) the models could not reproduce the stars with the highest
Rb abundances, which have [Rb/Fe] $\gtrsim$ 2.0.
A few possible solutions were suggested including extending
the calculations to include models of mass higher than 6$\Msun$.
Another solution was to consider variations of the mass-loss rate 
used on the AGB phase, which is known to be highly uncertain
\citep[e.g.,][]{groen09}.

The objective of our work is to explore two of the possible 
solutions to the models not producing enough Rb: 1) we extend our
calculations to include models of up to 9$\Msun$, and 2) we 
vary the AGB mass-loss rate in order to increase the number
of TPs and TDU mixing episodes. For the purposes of the
present study we concentrate on models of $Z=0.02$ (which in this
context are slightly super-solar metallicity, see \S\ref{newmodels}).
We expect that any of the proposed solutions, should they work, would apply 
equally well to stellar evolution models with metallicities appropriate
for stars in the Large and Small Magellanic Clouds (hereafter LMC and SMC).
This is because lower metallicity AGB stars already take many more
TPs to reach the superwind phase owing to their smaller radii and 
smaller pulsation periods than solar metallicity AGB stars.
We also examine the effect of variations in the nuclear network 
and variations in the \iso{22}Ne $+ \alpha$ reactions, including
adopting the latest rates for these reactions from \citet{iliadis10}.

The outline of this paper is as follows. We begin with a description 
of the new stellar evolutionary sequences and in 
\S\ref{nucleo} we present the new stellar nucleosynthesis predictions.
In \S\ref{sec:synthetic} we present the results of the synthetic 
evolution to the tip of the AGB and we finish with a discussion of
the new results obtained in \S\ref{sec:discussion}. 
Final remarks are given in \S\ref{sec:conclusion}.

\section{New Stellar Evolutionary Sequences} \label{newmodels}

We calculate the stellar evolution and nucleosynthesis in two steps. 
First, we use the stellar evolution code described in 
\citet[][and references therein]{karakas07b} 
to follow the evolution of the stellar structure and abundances important for stellar 
evolution (H, \iso{3}He, \iso{4}He, \iso{12}C, \iso{14}N, and \iso{16}O) 
from the zero-aged main sequence (ZAMS) to near the end of the AGB phase.
Then we perform detailed post-processing nucleosynthesis calculations to obtain 
heavy-element yields from the stellar evolutionary sequences.
In this study, we use the same stellar evolution code used in
\citet{karakas10b}, which includes the carbon and nitrogen-rich 
low-temperature opacity tables from \citet{lederer09}. The rate for
the \iso{14}N(p,$\gamma$)\iso{15}O reaction has been updated to
\citet{bemmerer06}, while the rate for the triple-$\alpha$ reaction
has been updated to NACRE \citep{angulo99}. 
We use the mixing length (MLT) theory for convective regions and we 
set the mixing length parameter $\alpha = 1.86$.
The temperature gradient in the convective envelope is determined by
the theory of convection used in the calculations. It can have a
substantial effect on the structure of the convective envelope and
influences surface properties such as the luminosity and effective
temperature. It also greatly influences the efficiency of HBB, with
the result that a shallower temperature gradient results in stronger
burning and higher luminosities \citep[see detailed discussion
in][]{ventura05a}.  In lower mass AGB stars, increasing $\alpha$ can
also increase the efficiency of the TDU \citep{boothroyd88d}.
In the intermediate-mass stars we are considering here,
the TDU is already very efficient ($\lambda \gtrsim 0.9$, see below) 
so it is not clear that increasing $\alpha$ will lead to a noticeable 
increase in the production of Rb. Regardless, we perform one 
calculation of a 6$\Msun$, $Z=0.02$ model with $\alpha = 2.5$.

For this study we concentrate on models with an initial composition 
of $Z=0.02$, where we set the solar metallicity to be $Z_{\odot} =
0.015$, very close to the proto-solar nebula solar metallicity of
0.0142 given in  \citet{asplund09}. The masses considered for this
study are $M = 5, 6, 7, 8$ and 9$\Msun$. For each of these masses 
we compute one evolutionary sequence using the standard 
\citet{vw93} mass-loss prescription. For the $M = 5, 6$ and $7\Msun$
models we compute one evolutionary sequence using the modified 
\citet{vw93} mass-loss prescription described below in \S\ref{sec:massloss}.
At this metallicity, we found that $M = 8\Msun$ is the limit 
for producing a C-O core, and even at 
this mass we found one carbon flash during which a small 
fraction of the C-O
core was burnt to neon \citep[similar to the models in][]{doherty10}.
The 9$\Msun$ model went through off-center carbon ignition and 
(almost) complete core carbon burning and entered the 
thermally-pulsing AGB with an O-Ne core.  AGB stars
with O-Ne cores are known as ``super-AGB stars'' and experience
HBB with temperatures at the base of the envelope of $T \gtrsim 
100 \times 10^{6}$~K. There are considerable uncertainties surrounding
the stellar modeling of super-AGB stars including the treatment of
core carbon ignition and burning and the efficiency of the
TDU \citep{siess10,doherty10,ventura11}. Here we simply use the
9$\Msun$ model as 
an illustrative example of the heavy-element nucleosynthesis in
super-AGB stars in comparison to lower-mass AGB stars with C-O cores.

\subsection{Mass loss in bright O-rich AGB stars} \label{sec:massloss}

Dealing with the extent and temporal distribution of 
mass-loss in AGB stars is a major uncertainty in stellar modeling
\citep{ventura05b}.  In the existing stellar evolution 
code, the \citet{reimers75} mass-loss prescription is used on the 
first giant branch with $\eta =0.4$, and the \citet{vw93} mass-loss
prescription is used on the AGB. 
According to this prescription the mass-loss rate depends
on the radial pulsation period and stays low ($\dot{M} \lesssim 
\times 10^{-6}\Msun$ year$^{-1}$) until the period, $P$,
exceeds 500 days. After this time a luminosity driven superwind 
begins and the mass-loss rate increases to a few $\times 
10^{-5} \Msun$ year$^{-1}$ and removes the envelope 
quickly, in a few TPs. 

\citet{vw93} noted that there are optically bright long period
variable stars with periods of $\approx 750$ days that are probably 
intermediate-mass stars of $\sim 5 \Msun$. In order to prevent their 
mass-loss prescription from removing these observed objects,
\citet{vw93} recommended a modification to delay the onset of the
superwind in stars of masses greater than 2.5$\Msun$. This 
suggestion is supported by the optical observations 
of OH/IR stars, which show that the number of obscured (dust 
enshrouded) stars dramatically increases for periods longer than 
700 days \citep[see Table 12 in][]{garcia07a}.
This modification was {\em not} used in the AGB model calculations
presented by \citet{karakas10a} and used in the \citet{vanraai12}
study.

Here we explore the effect of delaying the onset of
the superwind phase on the AGB nucleosynthesis. To do this, 
we adopt the modification to the \citet{vw93} mass-loss prescription. 
Specifically, we modify Equation~2 of \citet{vw93} such
that the superwind begins at a pulsation period of $P \approx 700-800$
days, depending on the initial stellar mass.
For comparison to observational data, the \citet{vw93} mass-loss
prescription gives pre-superwind mass-loss rates between
$\approx 10^{-8}$ to $\approx 10^{-6} \Msun$ year$^{-1}$, and superwind
mass-loss rates between a few $\times 10^{-5} \Msun$ year$^{-1}$
up to a maximum of $\approx 10^{-4}\Msun$ year$^{-1}$ in the most
massive AGB model of 9$\Msun$. These peak values are about a factor of
three to ten less than the maximum mass-loss rates measured in red
supergiants, which have mass-loss rates up to 
$\approx 10^{-3}\Msun$ year$^{-1}$ \citep{vanloon99b,debeck10}. 
\citet{debeck10} find a clear relation between the pulsation
periods of the AGB stars and the derived mass-loss rates, 
and note that the mass-loss rates level of at $3 \times 10^{-5}\Msun$ year$^{-1}$ 
for periods exceeding about 850 days. This suggests that delaying
the onset of the superwind to $P \approx 700-800$ days is a suitable
although somewhat conservative choice.

Here in this study we concentrate on variations to the \citet{vw93}
mass-loss prescription. There are several justifications for 
using the empirical mass-loss prescription from \citet{vw93} instead
of other empirical or theoretical mass-loss prescriptions
\citep[e.g.,][]{arndt97,bloecker95,wachter02,wachter08,vanloon05}. 
Importantly, the model stars we are considering are bright 
oxygen-rich AGB stars (i.e., C/O $<1$) for 
most of their TP-AGB phase owing to efficient HBB. This means that 
the mass-loss rates of e.g., \citet{arndt97} and \citet{wachter02} 
which were derived for C-rich stars, are not appropriate for our model stars. 
In comparison, the sample used by \citet{vw93} had Galactic and
Magellanic Cloud C-rich and O-rich AGB stars and covered a 
large range of pulsation periods and luminosities. 

Furthermore, the stellar effective temperatures of even the most
massive AGB model stars we compute are $\log T_{\rm eff} \approx
3.45$. These effective temperatures are similar 
to the M-type AGB stars in the Hertzsprung-Russell diagram of 
\citet[][their Fig. 12]{vanloon05}.
This means that M-type AGB mass-loss rates are more suitable for our model 
stars than red supergiant mass-loss rates, even though their predicted 
luminosities are near or just above the the classical AGB limit of
$\log L/\Lsun \approx 4.75$. Using the derived mass-loss rate for
M-type stars from \citet[][in particular, $\alpha =-5.65$, $\beta =
1.05$, and $\gamma = -6.3$]{vanloon05} and the typical stellar effective
temperature for the 9$\Msun$ star ($T_{\rm eff} \approx 2818$K) along
with the maximum AGB luminosity of $L \approx
80,000\Lsun$, we obtain a mass-loss rate of $\log \dot{M} = -4.11
\Msun$ year$^{-1}$. This is lower than the maximum mass-loss rate
of $\log \dot{M} = -3.96\Msun$ year$^{-1}$ found using \citet{vw93} 
for the 9$\Msun$ AGB model.  \citet{groen09}
compared the \citet{vw93} and Reimer's mass-loss prescriptions
to a sample of AGB mass-loss rates and concluded that the
\citet{vw93} still provides the best description. 
They did find however that there are also some deficiencies, 
in particular to the maximum adopted mass-loss rates. 

\subsection{Model results}

\begin{figure}
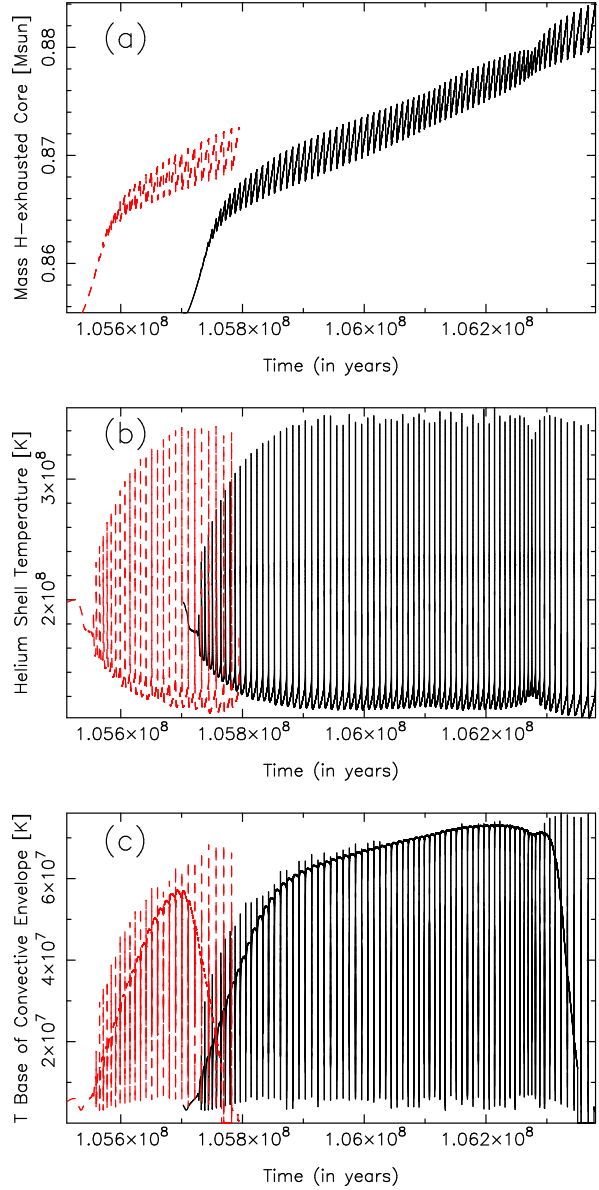

\begin{center}
\begin{tabular}{c}
\includegraphics[width=5cm,angle=270]{fig1a.ps} \\
\includegraphics[width=5cm,angle=270]{fig1b.ps} \\
\includegraphics[width=5cm,angle=270]{fig1c.ps} \\
\end{tabular}
\caption{The temporal evolution of the (a) growth of the
H-exhausted cores, (b) the He-shell temperatures, and (c) 
the temperatures at the base of the convective envelope
during the entire AGB. The black solid lines shows data for the
5$\Msun$, $Z=0.02$ model with a delayed superwind, and the red dashed
lines show data for the 5$\Msun$ with \citet{vw93} mass loss, shifted
to the left by $t = 1 \times 10^{5}$ years for clarity. 
(A color version of this figure is available on-line).}
\label{fig1}
\end{center}
\end{figure}

In Figs.~\ref{fig1} (a)--(c) we show the temporal evolution of 
(a) the H-exhausted core mass, the (b) temperature in the He-shell and
(c) the temperature at the base of the convective envelope for the
5$\Msun$, $Z=0.02$ models. The model with a delayed superwind is shown
by the black solid lines and  the model with the 
standard \citet{vw93} mass loss is shown by the red dashed lines 
(shifted to the left by $t = 1 \times 10^{5}$ years 
for clarity). The most striking feature of these figures is the
extra TPs in the model with a delayed superwind. This leads to 1) 
more TDU episodes and a final higher core mass, 2) more TPs where the
maximum temperature in the He-intershell exceeds $300 \times 10^{8}$K
and, 3) a longer duration for HBB and also a higher peak temperature,
which leads to a greater degree of proton-capture nucleosynthesis.
An examination of Table~\ref{table1} shows that the features
illustrated in Fig.~\ref{fig1} are present in all of the stellar 
models that delay the onset of the
superwind. The drastic increase in the number of TPs leads to 
increases in the total amount of He-intershell matter mixed into the 
envelope by the TDU for 5, 6, and 7$\Msun$ models. Furthermore, 
the peak temperature in the He-shell increases
with TP number and consequently the models with a delayed superwind
experience more TPs with temperatures sufficient to activate the
\iso{22}Ne neutron source,  with important consequences for Rb 
production as well as other heavy elements (see below).

In Table~\ref{table1} we show the results of the 6$\Msun$, $Z=0.02$
model with $\alpha=2.5$ and standard \citet{vw93} mass loss. The
strength of the TPs and the efficiency of the TDU episodes are almost the same
as the model with $\alpha=1.86$. Furthermore, the amount of He-shell
material dredged into the envelope over the AGB phase is almost identical
to the model with $\alpha = 1.86$. The most obvious difference is in 
the efficiency of HBB, which is more intense in the model with a higher 
$\alpha$. Given the similarities between the two models in terms of the 
amount of He-shell material and the strengths of TPs, we do not 
consider this model further in this study.

\begin{figure}
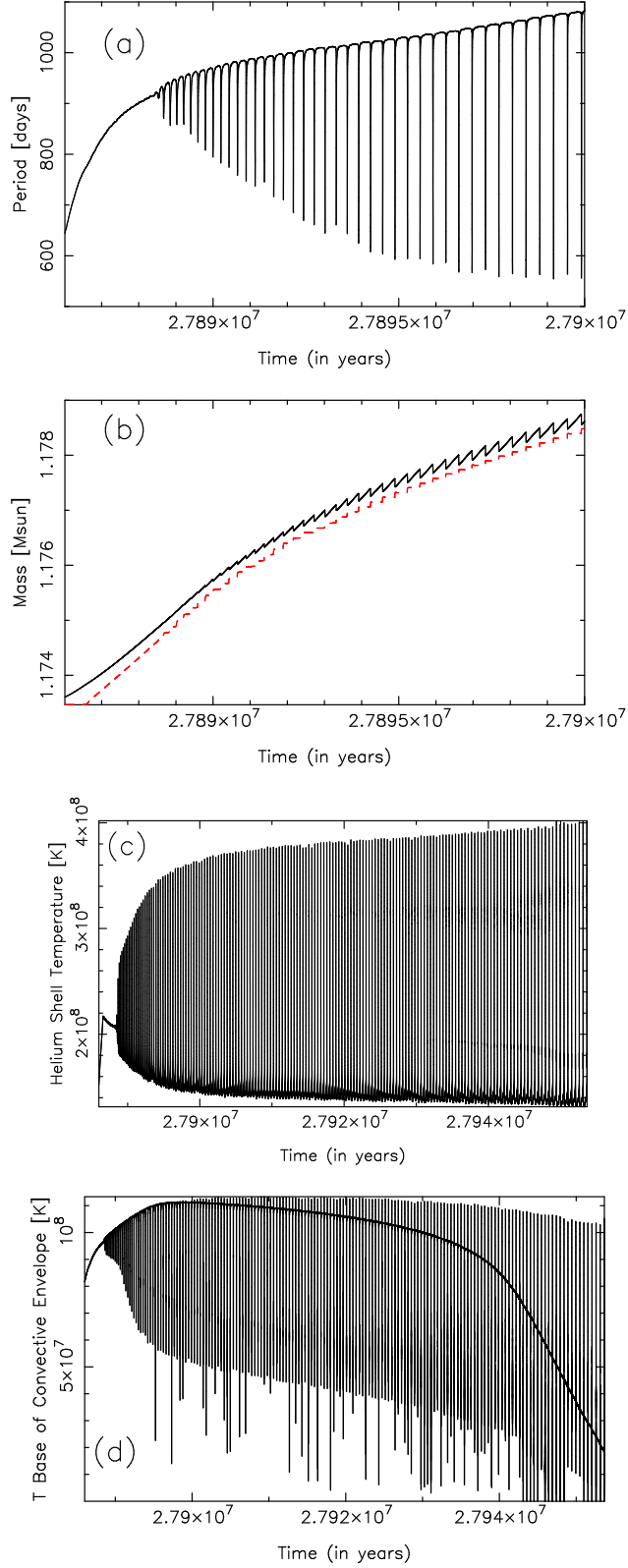

\begin{center}
\begin{tabular}{c}
\includegraphics[width=5cm,angle=270]{fig2a.ps} \\
\includegraphics[width=5cm,angle=270]{fig2b.ps} \\
\includegraphics[width=5cm,angle=270]{fig2c.ps} \\
\includegraphics[width=5cm,angle=270]{fig2d.ps}
\end{tabular}
\caption{The temporal evolution of the (a) pulsation period over the
first 40 TPs, (b) the growth of the H (black solid line) and
He-exhausted (red dashed line) cores over the first 40 TPs, (c) the He-shell temperature
over the entire AGB, and (d) the temperature at the
base of the convective envelope during the entire AGB for the 
9$\Msun$, $Z = 0.02$ model.}
\label{fig2}
\end{center}
\end{figure}

\begin{figure}
\begin{center}
\includegraphics[width=10cm,angle=270]{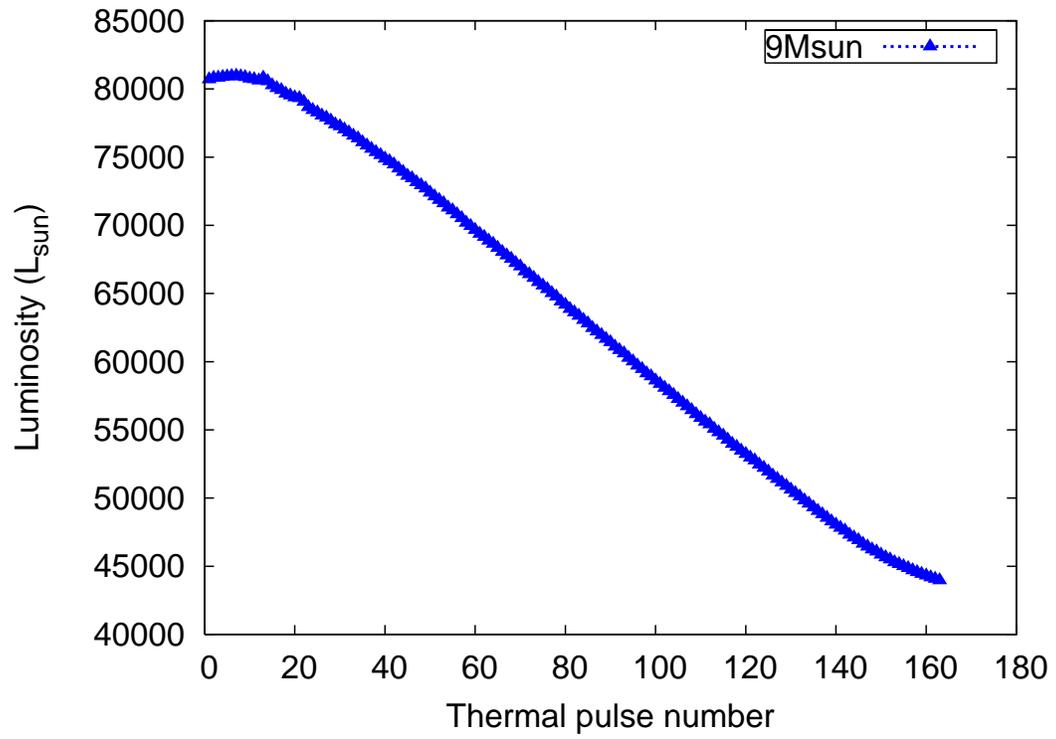} 
\caption{The evolution the stellar luminosity (in $L_{\odot}$) 
with TP number for the 9$\Msun$, $Z = 0.02$ model. (A color version 
of this figure is available on-line).}
\label{fig3}
\end{center}
\end{figure}

Both the 8$\Msun$ and 9$\Msun$ models enter
the thermally-pulsing AGB with pulsation periods $P \gtrsim 500$ days (see 
Fig.~\ref{fig2} (a) for the 9$\Msun$ model). This means that by the
first TP these stars are already experiencing the superwind and 
consequently the
behavior of the temperature at the base of the convective envelope 
reaches a peak early on and steadily declines with time (see
Fig.~\ref{fig2} (d) where the thick solid line represents the temperature
during the short interpulse periods). The peak in temperature at the 
base of the envelope is reached in $\approx 10$ TPs or so.
Similar behavior is noted for the evolution of the bolometric
luminosity. For example, in Fig.~\ref{fig3} we show the core-mass 
luminosity relation for the 9$\Msun$ model, where the luminosity
is observed to decrease monotonically with increasing core mass.
It is also instructive to examine the behavior of 
the TDU in these intermediate-mass models. The TDU efficiency
parameter $\lambda$ is shown to reach a peak at 5$\Msun$ and
then decrease with increasing stellar mass. This prediction
combined with the decreasing He-intershell mass means that even though
the 9$\Msun$ model experienced 163 TPs, the total amount of
He-intershell material dredged into the envelope is roughly a factor
of 2 lower than the 5$\Msun$ model with standard \citet{vw93} 
mass loss. However, the 9$\Msun$ model reaches peak He-shell temperatures of
$\approx 400 \times 10^{6}$K (Fig~\ref{fig2} (c)). This will have a
strong impact on the $s$-process nucleosynthesis, and in terms of the
final yields, will help to balance out the lower amount of 
dredged-up material.

\citet{siess10} presented the first stellar evolution and
nucleosynthesis predictions for super-AGB stars to the end of the
AGB phase. Our model compares reasonably well with that
of \citet{siess10} although the H-exhausted core mass of our model 
at the first TP is higher, at 1.17$\Msun$ compared to the same mass
and metallicity model computed by Siess (at 1.06$\Msun$) and that 
his model experiences 119 TPs compared to our 163 TPs. This could 
be simply a consequence of our model being evolved
to a lower final total mass of 3.2$\Msun$\footnote{Based on information 
from L. Siess's website, {\tt http://www.astro.ulb.ac.be/~siess/WWWTools/SAGB}
the final mass of his model was $M\approx 8.75\Msun$.}, at which point 
Fig.~\ref{fig2} (d) shows that HBB had ceased (the temperature at the
 base of the envelope is $\approx 2 \times 10^{7}$K).
Furthermore, our model experiences more efficient HBB with a peak
temperature of $111\times 10^{6}$ K compared to $99.2\times 10^{6}$ K
in the Siess model, and higher temperatures during He-shell burning
with a maximum of $402\times 10^{6}$ K compared to $341 \times 10^{6}$ K
in the Siess model. Finally, of consequence for the production of Rb
and other heavy elements, the standard models of \citet{siess10}
show no TDU while our 9$\Msun$ model does (see Fig.~\ref{fig2} (b)
which shows the evolution of the H-exhausted core). 
\citet{siess10} calculates a series of synthetic post-processing 
models where he sets the 
TDU efficiency parameter at a constant $\lambda = 0.8$. We will
compare to this model in \S\ref{nucleo}\footnote{Keeping in mind
that Siess's synthetic model does not take into account
the feedback of the TDU on the evolutionary properties of the 
model star.}.

Before we move onto discussing the results of the post-processing
nucleosynthesis results, we briefly mention a comparison of the new
5$\Msun$ and
6$\Msun$ models presented here with those presented in 
\citet{karakas10a} and used in
the previous \citet{vanraai12} study. The new stellar
evolutionary sequences were computed with updated C and N-rich
low-temperature opacity tables, which can have the effect of
increasing the radius and therefore the mass-loss rate when the star
becomes C and/or N-rich as a consequence of nucleosynthesis and
mixing \citep{marigo02}. Furthermore, the newer models use the slower
rate for the \iso{14}N(p,$\gamma$)\iso{15}O reaction from \citet{bemmerer06}, 
which effects the rate of H-burning during the interpulse phase.
For most of the masses considered here,
efficient HBB prevents the formation of a C-rich atmosphere so the
question of C enrichment should not be important, however all models
become N-rich. At this metallicity, 5$\Msun$ is the mass at which HBB
begins and \citet{ventura09b} showed that the 
stellar yields of the lowest mass models that experience HBB crucially
depend on the adopted molecular opacities \citep[see also][]{ventura10}.
The  new 5$\Msun$ model is similar to the previous model
(similar number of TPs, peak HBB temperatures and luminosities) but dredged up
$\approx 30$\% more He-intershell material as a consequence of the 
new model being evolved to a smaller envelope mass. The new
6$\Msun$ reveals a larger deviation in that it has 15\% fewer
TP (33TP compared to 38TP) and lower peak interpulse luminosities 
and HBB temperatures by roughly 20\% and 10\%, respectively.
This is in spite of the higher mixing-length parameter of 1.86
in the new models compared to 1.75 previously \citep[][showed that
higher $\alpha$ leads to higher luminosities and
temperatures]{ventura05a}. The lower temperatures and luminosities
is likely caused by using the lower \iso{14}N(p,$\gamma$)\iso{15}O
rate. Overall though, important for $s$-process nucleosynthesis 
is the amount of material dredged to the stellar surface. 
Even though the new model experienced fewer TPs it was evolved to
a lower envelope mass and dredged up 13\% more He-shell material
compared to the old model. In summary the newer models, while 
computed with improved input physics, show results that are 
reasonably consistent with the previous calculations. 

%

\section{Nucleosynthesis Results} \label{nucleo}

The stellar evolutionary sequences described above are fed into a
post-processing code in order to obtain nucleosynthesis predictions
for elements heavier than iron. The details of this procedure have
been described in detail elsewhere including \citet{karakas09} and \citet{lugaro12}.
Briefly, the nucleosynthesis code needs as input from the stellar
evolution code variables such as temperature, density, and convective 
boundaries as a function of time and interior mass. The code then 
re-calculates the abundance changes as a function of mass and time 
using a nuclear network which contains $N$ species and time dependent 
diffusive mixing for all convective zones \citep{cannon93}. For this
study we utilize three nuclear networks: 1) 166 species using the 
same rates used in the study of \citet{vanraai12}, 2) 172 species and
based on the latest JINA REACLIB database \citep{cyburt10}, and 3) a
320 species network for a selection of models to examine predictions
for hydrogen through to lead. The 166 and 172 species networks include
elements from hydrogen to sulfur, and then from Fe to Nb (166) or Mo
(172) and includes elements of the first $s$-process
peak at Rb, Sr, Y, and Zr.

The production of elements heavier than Fe in intermediate-mass 
AGB stars crucially depends on the rate of the
\iso{22}Ne($\alpha$,n)\iso{25}Mg reaction. In the 166 species network
we use the \citet{karakas06a} rate used by \citet{vanraai12}, 
whereas in the 172 and 320 species network we
use the NACRE \citep{angulo99} rate. As a further test, we perform
calculations using the 172 species network and use the latest rates
of the \iso{22}Ne($\alpha$,n)\iso{25}Mg and \iso{22}Ne($\alpha,\gamma$)\iso{26}Mg
reactions from \citet{iliadis10}.

For the initial composition we took the solar distribution of 
abundances from \citet{asplund09} which sets the global solar
metallicity to be $Z_{\odot} = 0.0142$ which we round up to 0.015.
We scale the solar abundances to a metallicity of $Z=0.02$ which
results in a slightly super-solar initial [Fe/H] of 0.14.
Solar abundances of C, N, O, Ne, Mg, Si, S, Ar, and Fe are the 
pre-solar nebula values from Table~5 of \citet{asplund09}; F is 
the meteoritic value of $\log \epsilon$ (F)$_{\odot}$ = 4.42 
from Table~1 of the same paper, chosen because it has a lower uncertainty.
For the same reason we use the meteoritic values for the 
solar abundances for many elements heavier than Fe including
Sr, Eu, and Pb. 

\begin{figure}
\begin{center}
\includegraphics[width=10cm,angle=270]{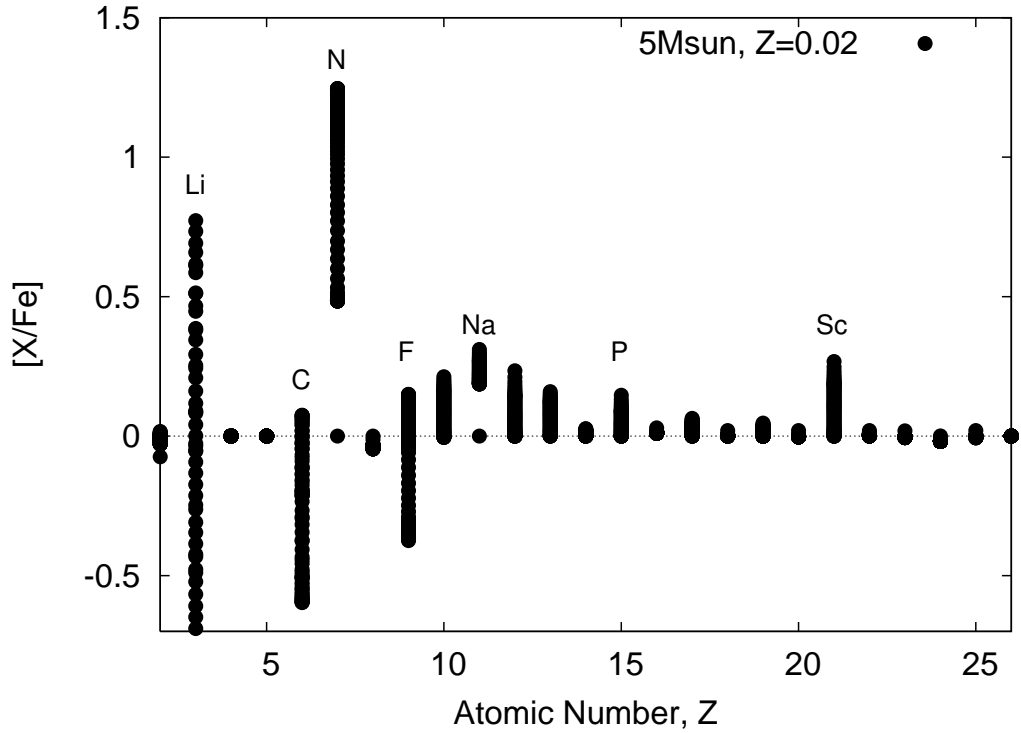} 
\caption{The evolution of elements [X/Fe] lighter than Fe 
as a function of atomic number, $Z$,
for the 5$\Msun$ model with a delayed superwind and nuclear network
using 320 species. Included are the approximate locations (in proton
number, $Z$) of some key elements. Each dot represents the surface
composition after a thermal pulse.}
\label{fig4}
\end{center}
\end{figure}

\begin{figure}
\begin{center}
\includegraphics[width=10cm,angle=270]{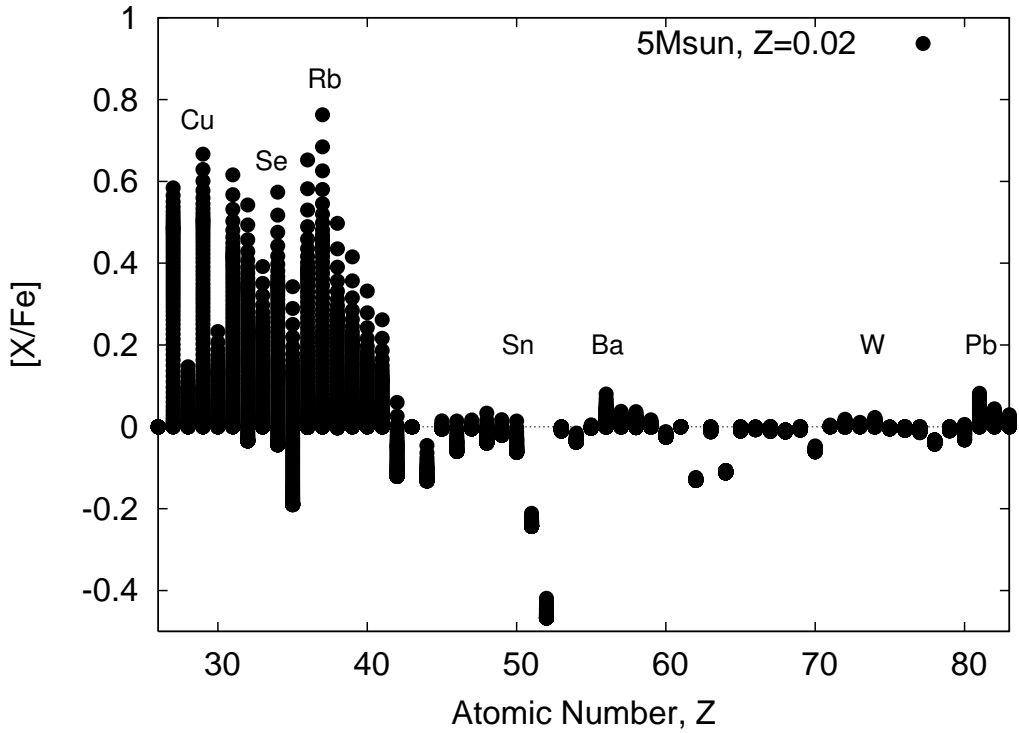} 
\caption{The evolution of elements [X/Fe] heavier than Fe as a 
function of atomic number, $Z$, for the 5$\Msun$ model with a 
delayed superwind and nuclear network using 320 species. Included are
the approximate locations of some key elements, noting that Pb 
is at $Z=82$. Each dot represents the surface
composition after a thermal pulse.}
\label{fig5}
\end{center}
\end{figure}

\begin{figure}
\begin{center}
\includegraphics[width=10cm,angle=270]{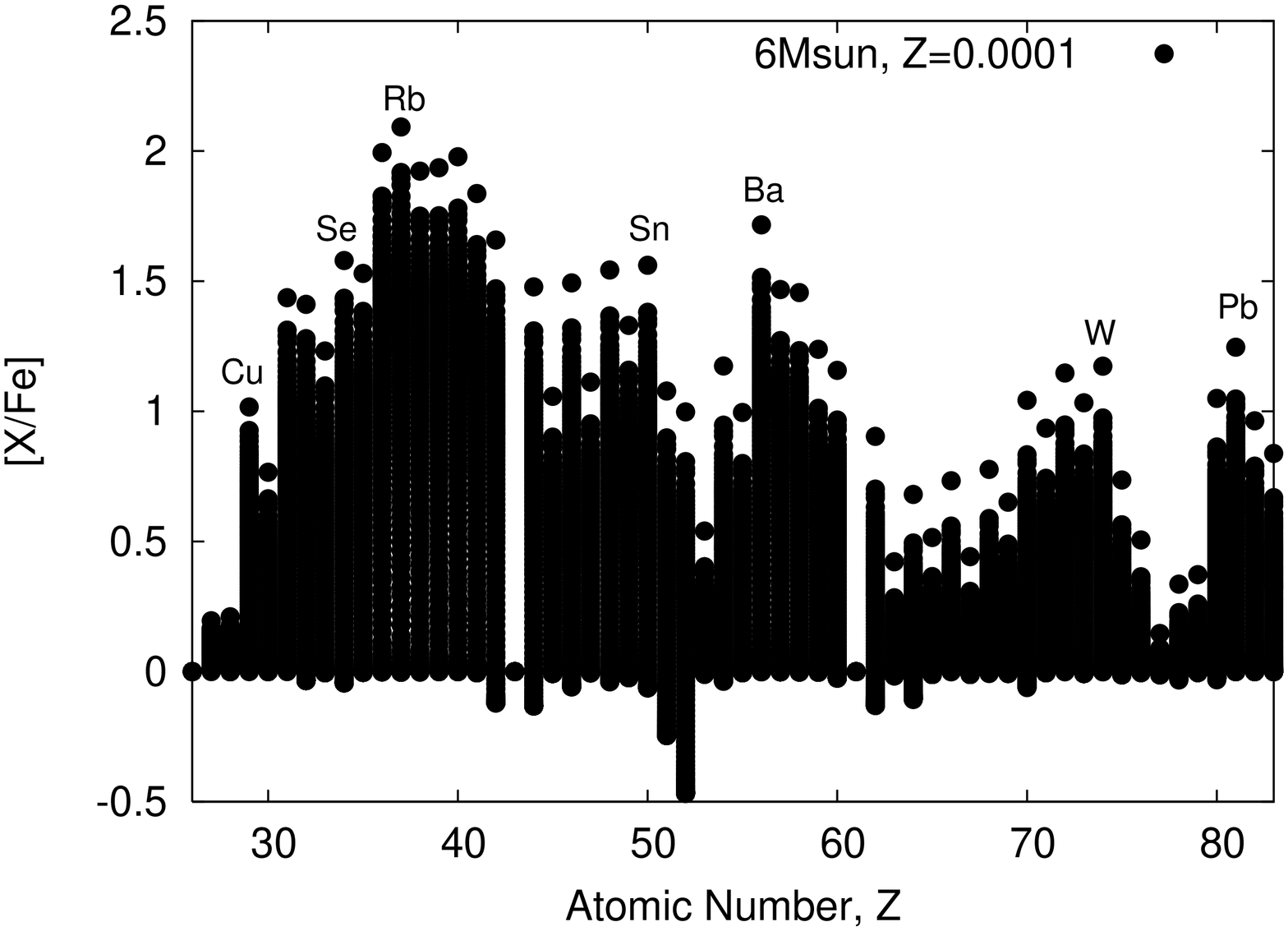} 
\caption{The evolution of elements [X/Fe] heavier than Fe as a 
function of atomic number, $Z$, for the 6$\Msun$, $Z=0.0001$ model 
with scaled solar initial abundances from \citet{lugaro12}. 
The nuclear network used for this calculation had 320 species
and included the \iso{22}Ne + $\alpha$ rates from \citet{karakas06a}.
Included are the approximate locations of some key elements, 
noting that Pb is at $Z=82$. Each dot represents the surface
composition after a thermal pulse.}
\label{fig6}
\end{center}
\end{figure}

In Figs.~\ref{fig4} and~\ref{fig5} we show the nucleosynthetic
results for the elements lighter and heavier than Fe, respectively, 
from the  5$\Msun$ model with a  delayed superwind and 320 species.
For some of the heavier elements, the initial [X/Fe] $<0$ (e.g.,
Sb and Te) because we do not include all of the stable 
isotopes in the network. For example for Te, a large fraction 
(roughly 65\%) of the elemental component comes from
isotopes produced by the rapid neutron capture process (i.e.,
\iso{128}Te and \iso{130}Te) and we do not include these isotopes in
our 320 species network. Overall, the nucleosynthesis is as expected
with significant N and Rb production as well as some surprises
including some Sc production ([Sc/Fe] $\approx 0.3$~dex), along
with Co and Cu production \citep[at the level of 0.6~dex for Co and
Cu; see also][]{smith87}. Carbon shows a strong variation with a decrease down to [C/Fe] 
$\approx -0.6$ followed by an increase to $\approx 0.07$ by the tip of the AGB.
Also from Fig.~\ref{fig5} we see no significant production of elements
beyond the first $s$-process peak. This is not the case as
the metallicity is reduced: the $s$-process predictions for the 
6$\Msun$, $Z=0.0001$ model presented by \citet{lugaro12} and plotted
in Fig.~\ref{fig6} shows instead significant Ba and Pb production (final
[Ba/Fe] = 1.72 and [Pb/Fe] = 0.96). The low-metallicity 6$\Msun$ model 
also produces some Sc ([Sc/Fe] = 0.47) and Cu ([Cu/Fe] = 1.0) although
little Co. Note that the 6$\Msun$, $Z=0.0001$ model predictions
illustrated in Fig.~\ref{fig6} would be even more extreme if we
were to use the NACRE rate for the \iso{22}Ne($\alpha$,n)\iso{25}Mg 
reaction.

In Tables~\ref{table2} and~\ref{table3} we present a selection of 
elemental and isotopic abundance ratios at the tip of the AGB for the
nucleosynthesis models computed for this study. In each case we list
the mass,  the pulsation period ($P$) at which the superwind begins,
the network used for the calculation, the rate for the
\iso{22}Ne($\alpha$,n)\iso{25}Mg reaction, and then a selection of
isotopic and elemental abundance ratios, where we adopt the standard
spectroscopic notation, [X/Fe]. All isotopic and element ratios were
computed using abundances by number. We find good agreement between
models using the 172 and 320 species networks and the NACRE rate for the
\iso{22}Ne neutron source. Table~\ref{table2} shows that the 
production of elements beyond the first $s$-process peak (Ba as an
example) is negligible and hence our assumption of neglecting these
species is valid for this study. Now we compare results from models 
computed using the 172 species network but different rates for the
\iso{22}Ne $+ \alpha$ rates \citep[from NACRE and from][]{iliadis10}.
Using the NACRE rate results in the highest levels of Rb and 
neutron-capture element production although there is only 
$\approx 0.1-0.3$~dex difference between calculations that use the
NACRE rate compared to those that use the \citet{iliadis10} rate. 
The 166 species
network used in the \citet{vanraai12} study results in the lowest
levels of neutron-capture nucleosynthesis. There are other 
differences between the two networks besides the rate of the
\iso{22}Ne $+ \alpha$ reactions. The 166 network includes
older proton capture rates for the CNO cycle and the NeNa and
MgAl chains, along with fits to the neutron-capture cross 
sections of \citet{bao00} whereas the updated 172 network includes 
JINA REACLIB fits to the KADoNiS 
database\footnote{Website: http://www.kadonis.org}.


In terms of the nucleosynthesis of elements lighter than Fe, we find
the effects of HBB are clearly evident in all models, including the
5$\Msun$ model with a standard \citet{vw93} mass loss which becomes
C-rich at the final TPs. All models show enhanced lithium, nitrogen and sodium,
mild to moderate oxygen destruction, and small increases in magnesium and
aluminum. The observations of \citet{garcia07a} show that some stars 
enhanced in Rb have high Li and some have low Li \citep[see
discussion in][]{vanraai12}. The final Li abundances of our models
also show a large variation where $\log
\epsilon$(Li)\footnote{Defined by $\log \epsilon$(Li) = $\log$(Li/H)
$+ 12$ and the abundances are by number.} varies between 2.4 for the
5$\Msun$ model with a delayed superwind to $-2.2$ for the 7$\Msun$
with a delayed superwind. The 8$\Msun$ and 9$\Msun$ models 
have final $\log \epsilon$(Li) = 2.0 and 1.9, respectively. 

In all cases the Mg isotopic ratios are also altered from
their initial solar values, and most of all in the 9$\Msun$ model with
the most extreme HBB. In this case the alteration of the Mg isotopes
is mostly the result of proton-captures via the Mg-Al chains. 
For the lighter elements we find good agreement between all
nuclear networks for C, N, and O. The [Na/Fe] ratio is shown to be
significantly lower in the updated 172 and 320 species networks; this
is a result of using the newer \iso{23}Na $+$ p rates from
\citet{hale04} whereas the 166 species network uses the NACRE rates
\citep[see discussion in][]{karakas10a}.
The elemental [Al/Fe] predictions are reasonably
consistent between the different networks at a given mass, although
the effect of using different \iso{22}Ne $+ \alpha$ rates is evident
in the predictions for the Mg isotopic ratios. 

In \citet{karakas10a} we compared our nucleosynthesis results for AGB
stars of $\le 6\Msun$  to other groups. Given the
consistency between the older models and those presented here 
we do not repeat that discussion. \citet{siess10}
provides yields -- in terms of the mass (in $\Msun$) of each 
element ejected by the stellar winds -- for the super-AGB stars
considered in his study. Here we compare the
results from Siess's synthetic 9$\Msun$, $Z=0.02$ nucleosynthesis model 
with a constant $\lambda = 0.8$ and 342 TPs to our 9$\Msun$ model, 
which has $\lambda \approx 0.77$ and 163 TPs.
For this comparison we use yields from the 172 species network.
The hydrogen yield is essentially the same in our calculations, 
our \iso{4}He yield is only 3\% lower (where the final Y $\approx
0.38$), our \iso{12}C yield is about 60\% lower although our yield 
of \iso{14}N is 3\% higher, indicating more processing by HBB.  
We also include a synthetic extension for this model 
(see \S\ref{sec:synthetic}) that results in an extra 60 TPs 
bringing the total to 223 TPs and increases in \iso{12}C 
but no further increase in \iso{14}N because HBB had ceased. 
Keeping in mind the qualitative nature of the yields presented 
by \citet{siess10} and that our model experienced many fewer
TPs, these differences are relatively small. 

The study by \citet{vanraai12} investigated the effect of including
\iso{13}C-rich regions (``pockets'') in the He intershell of 
intermediate-mass AGB stars. It is in the \iso{13}C pocket that
neutrons are produced via the alternative neutron source, 
\iso{13}C($\alpha$,n)\iso{16}O. \citet{vanraai12} showed that including 
\iso{13}C pockets increase the production of Rb by up to 0.4~dex. 
In this study we perform one test calculation where we include some extra
mixing of protons into the He-intershell of the 5$\Msun$ 
$Z =0.02$ model with a delayed superwind. We do this in the same
manner as described by \citet{vanraai12} and \citet{kamath12}. 
We assume that protons are partially mixed into the intershell over a region of mass, 
$M_{\rm mix}$, at the deepest extent of each TDU episode, where the 
number of protons decreases exponentially from the envelope composition, 
$X_{\rm p} \approx 0.7$, to  $X_{\rm p} = 1\times 10^{-4}$. In this test
case we set $M_{\rm mix} = 1\times 10^{-4}\Msun$ for the 5$\Msun$
model, where $M_{\rm mix}$ is constant. In general, the mass of 
the He-intershell is about a factor of 10 larger than $M_{\rm mix}$.
The final surface [Rb/Fe] is 0.93~dex, an increase of about 0.2~dex
on the model without a  \iso{13}C pocket. This is smaller than
found by \citet{vanraai12} at the same mass and metallicity, although
we note that the 5$\Msun$, $Z = 0.02$ model used by \citet{vanraai12}
with \citet{vw93} mass loss only produced a final [Rb/Fe] = 0.05~dex.
These results are encouraging and suggest that the inclusion of 
a small \iso{13}C pocket could lead to [Rb/Fe] ratios consistent with
the average of the observed abundances, when considering models with
a delayed superwind.

\section{Synthetic evolution to the tip of the AGB} \label{sec:synthetic}

In all cases the stellar evolutionary sequences were not evolved until
the envelope mass was removed and the model star had left the AGB.
This is a result of convergence difficulties that arise once the
envelope mass drops below some value ($M_{\rm env} \approx 1\Msun$ depending on
initial mass, see Table~\ref{table1}). In all cases convergence
difficulties end the calculation after the cessation of HBB 
(e.g., see Figs.\ref{fig1} and \ref{fig2}). The distribution of
white dwarf (WD) masses reveals the existence of massive WDs 
($\approx 0.8\Msun$ to 1.2$\Msun$) originating from single star evolution
\citep{liebert05,ferrario05}.  This suggests that
whatever the cause of the convergence difficulties, real stars somehow
lose their envelope at the tip of the AGB. The real uncertainty
is whether or not the model stars should experience further TPs and
TDU episodes near the tip of the AGB \citep{lau12}.

Assuming that TPs and TDU episodes will occur, it is possible to
estimate the effect of these extra TPs on the surface compositions of 
our model stars. We do this in the same manner described by
e.g., \citet{karakas07b} and \citet{vanraai12}.
Our method is simple by synthetic AGB evolution standards
\citep{marigo01,izzard04b} although of a similar complexity to 
the synthetic extensions to the nucleosynthesis used by e.g.,
\citet{forestini97} for AGB stars, and \citet{siess10} for super-AGB
stars. For further details we refer to the discussion
in \citet{karakas07b}.
Our method assumes that all of the physical quantities of the star
(e.g., luminosity, interpulse period, TDU efficiency etc)
are assumed to remain constant at the  tip of the AGB. The study by
\citet{karakas07b} showed that this is a reasonable assumption at the
tip of the AGB in intermediate-mass stellar models. The main 
uncertainty of this method is the efficiency of the TDU, which we
assume remains the same throughout our synthetic calculations. 
Theoretical models of intermediate-mass stars with deep TDU find
that the efficiency does not vary significantly, even with decreasing
envelope mass \citep{karakas10a}.
For a selection of stellar evolutionary sequences we estimate the 
number of remaining TPs; this is
shown in Table~\ref{table4}. The stellar mass and pulsation period ($P$) at
which the superwind begins are listed together with the amount 
of material lost at each interpulse period ($\Delta M$), the Rb 
intershell mass fraction (X(Rb)$_{\rm intershell}$), and the 
final [Rb/Fe] ratio predicted by our synthetic calculations.  
In all cases we take intershell compositions from the 172 species
network with the NACRE rate for the \iso{22}Ne reaction, as these
should provide an upper limit to the final [Rb/Fe] ratio. 

For the 9$\Msun$ model the synthetic algorithm estimated that there 
should be 60 TPs remaining and the final [Rb/Fe] = 0.90.
For this model, the large number of synthetic TPs stretches our
assumption of a constant luminosity, interpulse period, intershell
composition, etc. (e.g., see Fig.~\ref{fig3} which shows the 
evolution of the luminosity with core mass). Varying the 
interpulse period and luminosity by reasonable amounts ($\approx 20$\%
over 60 TPs) changes the final estimate for [Rb/Fe]
by only $\approx 0.05$~dex. The largest change to the final
[Rb/Fe] ratio is obtained by increasing the intershell composition of Rb.
Over 60 TPs, the intershell composition of Rb increased by about
a factor of two, as a result of TPs becoming hotter with evolution.
Assuming that the Rb intershell mass fraction increases again
by a factor of two we obtain a final [Rb/Fe] of 1.12~dex, which
we consider as the maximum value for this particular model.
Finally, the synthetic evolution algorithm results in a final 
H-exhausted core mass of $1.19\Msun$, still well below the 
critical core mass value of $\approx 1.37\Msun$ \citep{nomoto84}, 
above which electron capture reactions start and the core collapses. 
This is also the case if we were to assume that there are no more TDU
episodes, where we predict a final core mass of 1.22$\Msun$.

The main point of this section is to highlight that it is possible
to reach [Rb/Fe] values of $\gtrsim 1.3$~dex in the 6$\Msun$ and
7$\Msun$ models simply by extending the evolution and allowing for
TPs near the tip of the AGB.
In all of the models, the synthetic extension causes the 
surface composition to become carbon rich, where C/O $> 1$, 
owing to the fact that HBB has ceased but the TDU continues.
For this reason we conclude that TPs that occur after the cessation
of HBB are not the cause of the highest Rb overabundances observed
in the OH/IR stars. 

\section{Discussion} \label{sec:discussion}

\begin{figure}
\begin{center}
\includegraphics[width=10cm,angle=270]{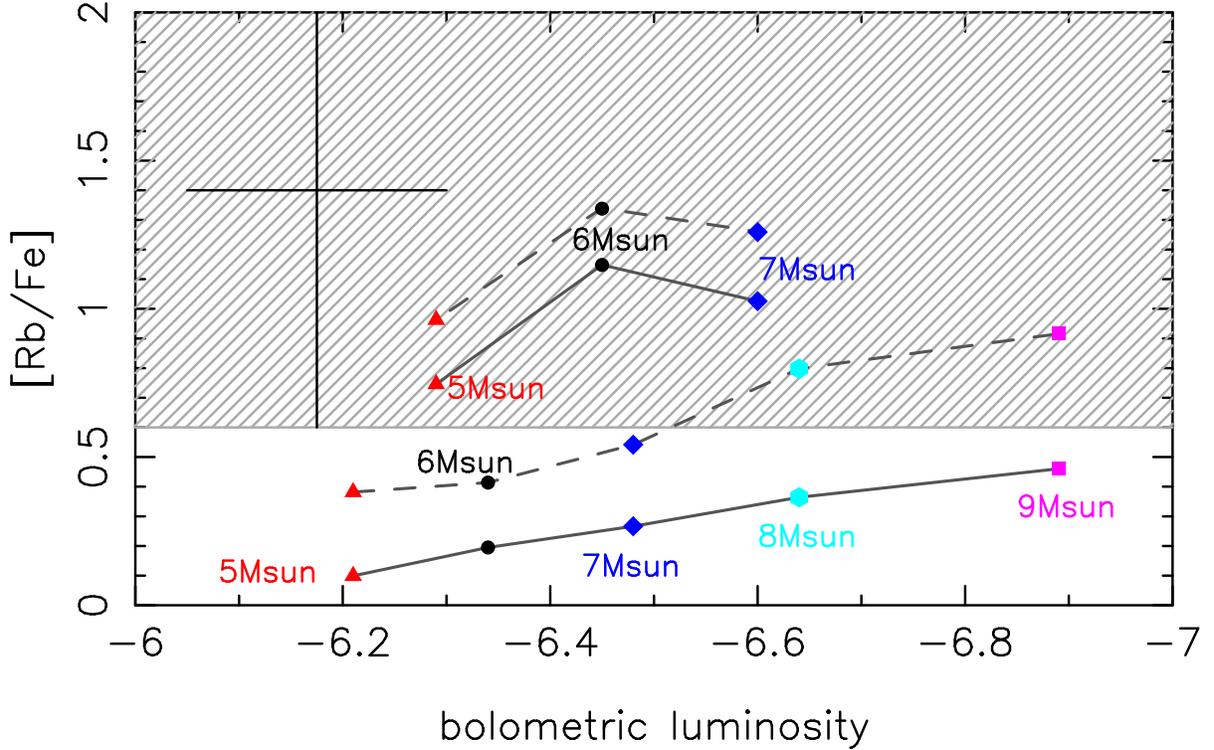} 
\caption{The bolometric luminosity at the tip of the AGB versus the
[Rb/Fe] abundance from the last computed TP (connected by 
the solid lines) and from the synthetic evolution calculations 
(connected by the dashed lines). Models using the 
\citet{vw93} mass-loss prescription are connected by the lower
solid and dashed lines, and models calculated using a delayed
superwind by the upper solid and dashed lines, respectively.
Symbols and labels denote the initial stellar mass.  The shaded
region indicates the range of observed [Rb/Fe], noting that 
the maximum reaches $+$2.5 dex. The average observed [Rb/Fe] = 1.4 is shown
on the left along with the maximum uncertainty of $\pm$0.8~dex.
(A color version of this figure is available on-line).}
\label{fig7}
\end{center}
\end{figure}

First we examine the behavior of [Rb/Fe] as a function of mass (or
bolometric luminosity) for the set of models computed using the 
\citet{vw93} mass-loss prescription.
In Fig.~\ref{fig7} we show the bolometric luminosity at the tip of the
AGB versus the [Rb/Fe] ratios for models calculated using the \citet{vw93}
mass-loss formula and for models with a delayed superwind.
We show results from the 172 species network that used the NACRE 
rate for the \iso{22}Ne reaction, and include the final computed 
and synthetic [Rb/Fe] from Tables~\ref{table3} 
and~\ref{table4}, respectively. The average [Rb/Fe] of 1.4~dex
of the Rb-rich stars is included along with the maximum uncertainty
on the derived abundances. The dashed region corresponds to the range
of [Rb/Fe] where Rb-rich stars are found, where we choose a range
of 0.6~dex to 2~dex \citep[note that][choose a lower limit of
0.4~dex]{vanraai12}. We can see that the most massive 
AGB stars produce the most Rb, as suggested by the observations 
of \citet{garcia06} and
found theoretically for a smaller mass range by \citet{vanraai12}. 
This trend holds when we compare [Rb/Fe] values obtained at  the final
TP (lower solid line in Fig.~\ref{fig7}) compared  to results
from the synthetic extension (lower dashed line in Fig.~\ref{fig7}). 
From inspection of Fig. 2 of \citet{garcia06} we see that most 
(12) of Rb-enriched Galactic 
OH/IR stars have [Rb/Fe] between $+0.6$ dex and $+1.6$ dex (with an 
average of $+1.4$ dex), with a maximum error of $\pm 0.8$ dex. 
Three stars have [Rb/Fe] ratios much higher, at 
[Rb/Fe] $\gtrsim 2 \pm 0.8$ dex.
None of  the models connected by the lower solid line in Fig.~\ref{fig7}
produce [Rb/Fe] ratios in the range of the observations, and only the
synthetic extension can produce high enough [Rb/Fe] ratios in the
8$\Msun$ and 9$\Msun$ models. 

From Fig.~\ref{fig7} we see that models with a delayed superwind
produce higher [Rb/Fe] abundances (shown by the upper solid line) 
and are found within the shaded region of the observations.
This is mostly because these models experience many more TPs and TDU 
episodes. For example, delaying the superwind results in maximum
increases of $\Delta$[Rb/Fe] = [Rb/Fe]$_{\rm dsw}$ - [Rb/Fe]$_{\rm
vw93}$ = 0.65, 0.95, and 0.76~dex respectively for the 5, 6, and 
7$\Msun$ models when comparing models that use the NACRE rate.  
Note that [Rb/Fe]$_{\rm dsw}$ is the final surface [Rb/Fe]
from the model with a delayed superwind and [Rb/Fe]$_{\rm vw93}$ the
final [Rb/Fe] from the model with \citet{vw93}. We find smaller 
increases of $\Delta$[Rb/Fe] = 0.25, 0.58, and 0.50~dex when 
using the 166 species network and the rate from \citet{karakas06a}. 
In comparison, we find smaller
variations if we compare nucleosynthesis models computed with 
the same stellar evolution input but different nuclear networks 
and rates for the \iso{22}Ne $+ \alpha$ reactions. For example,
using the newer \citet{iliadis10} rates for
the \iso{22}Ne $+ \alpha$ reactions instead of NACRE in the
models with a delayed superwind results in decreases of up to 0.3~dex
in the final [Rb/Fe] abundance.  There are larger variations of up to 
0.5~dex between the 172 species network delayed superwind models 
with the NACRE rate and models computed with the 166 species 
network with the \citet{karakas06a} rate. \citet{vanraai12} did 
not find such large variations and indeed they only show up for 
the models with a delayed superwind.  This is a consequence of 
the models experiencing many more TPs at temperatures that allow 
for activation of the \iso{22}Ne neutron source.

\begin{figure}
\begin{center}
\includegraphics[width=10cm,angle=270]{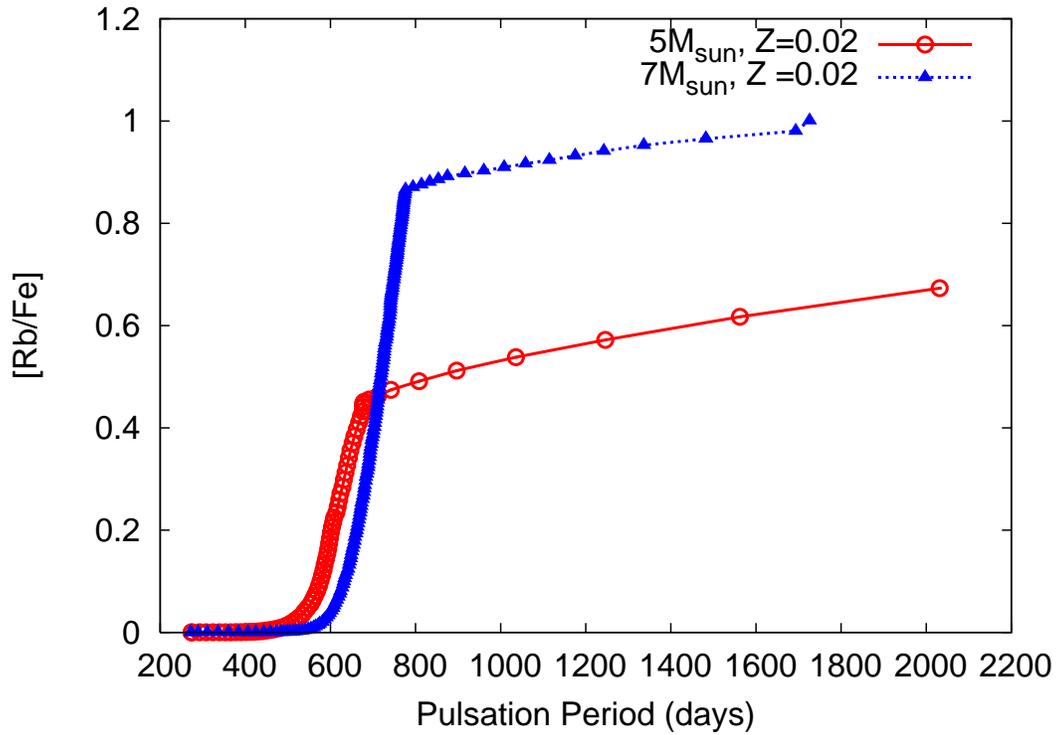} 
\caption{The pulsation period from the stellar evolutionary sequence
versus [Rb/Fe] from the post-processing nucleosynthesis calculations
for the 5$\Msun$ and 7$\Msun$ models with a delayed superwind.
(A color version of this figure is available on-line).}
\label{fig8}
\end{center}
\end{figure}

In Fig.~\ref{fig8} we show the surface [Rb/Fe] abundance versus
the pulsation period for two models with a delayed superwind. 
Here the [Rb/Fe] value is seen to increase reasonably quickly 
for pulsation periods higher than about 500 days and 600 days
for the 5 and 7$\Msun$ models, respectively. Once the superwind
starts, the [Rb/Fe] abundance is seen to plateau. 
This is because the [Rb/Fe] does not increase very much from
one TP to another ([Rb/Fe] only increases by
$\approx  3-10\times 10^{-3}$~dex per TP). During the superwind the
pulsation period increases quickly but [Rb/Fe] does not.
The predicted [Rb/Fe] of the 7$\Msun$ model enters the 
lowest observed abundance range (0.6~dex) of the Rb-rich stars
at a pulsation period of approximately 700 days, whereas the 
5$\Msun$ only does so near the end of the calculation.
The observed period versus [Rb/Fe] relationship for the Rb-rich OH/IR
stars is not published but a comparison would prove useful for
providing extra constraints on the stellar evolution calculations.


Models with a delayed superwind also produce 
the highest Rb intershell abundances, again a consequence of more
TPs. An increase in TPs leads to an increase in core mass, which
leads to an increase in the maximum He-shell
temperature. An increase in temperature results in
a stronger activation of the \iso{22}Ne($\alpha$,n)\iso{25}Mg reaction
and more Rb production. However it is instructive to compare 
the results for the 9$\Msun$ super-AGB model in comparison to the 6 and 
7$\Msun$ models computed with a delayed superwind. These models
experienced similar numbers of TPs with temperatures over $300 \times
10^{6}$~K (150, 178, 155 TPs for the 6, 7, and 9$\Msun$ models
respectively) but the 6 and 7$\Msun$ models produce more Rb in the
intershell. To understand why we need to examine Rb production
in more detail. \citet{vanraai12} provided a detailed discussion
of the neutron-capture pathways that lead to the production
of Rb in intermediate-mass AGB stars, here we briefly recap
how this happens. 
Elemental Rb consists of two stable isotopes, \iso{85}Rb 
and \iso{87}Rb.  The neutron-rich isotope \iso{87}Rb is produced
in almost equal quantities in He-intershells of the 6$\Msun$ and 
9$\Msun$ models (at the last TP the \iso{87}Rb abundance is only
about 10\% higher in the 6$\Msun$ model). However, in the 9$\Msun$
model the \iso{85}Rb/\iso{87}Rb ratio in the He-intershell 
after the last TP is 0.15 and therefore dominated by \iso{87}Rb.
In comparison  the ratio is 0.42 in the 6$\Msun$ model after the last 
TP. For comparison, the solar \iso{85}Rb/\iso{87} ratio $= 2.43$
\citep{asplund09}. This indicates that \iso{85}Rb is not being
produced as efficiently in the 9$\Msun$ model. 
This is because most of the \iso{85}Rb comes from the decay of the
ground-state of the unstable \iso{85}Kr, which has a half-life of
3939 days (or $\approx 10.7$ years)\footnote{In our networks we also include
as a separate species the meta-stable state of \iso{85}Kr which has a
half-life of 4.48 hours and also $\beta$-decays to \iso{85}Rb.}.
This lifetime of 10.7 years is longer than the duration of 
convective TPs ($\tau \lesssim 3$~years) in the 9$\Msun$ model,
and also less than duration of the TDU ($\tau \lesssim 3$~years).  
This means that most of the \iso{85}Kr decays to \iso{85}Rb after
mixing has finished. 
Note that the next TP is ignited further out in the
He-intershell, leaving the previously made \iso{85}Rb in the core.
In contrast, the duration of convective TPs in the 6$\Msun$ model is
about 15 years, which allows the \iso{85}Kr to decay before the onset
of the subsequent TDU episode.

We can now examine if either of the two proposed solutions we
are testing can produce enough Rb to match the observations. Increasing
the stellar mass does produce more Rb and leads to higher [Rb/Fe]
ratios. The 8 and 9$\Msun$ models can produce enough Rb to match the
bulk of the Rb-rich stars with [Rb/Fe] $\approx 1$~dex but only when
synthetically extending the calculation to account for remaining TPs.
Given the uncertainty of the synthetic calculation for the 9$\Msun$
model we take this result as tentative.
Other calculations of $s$-process nucleosynthesis of AGB stars near 
the C-O core limit as well as super-AGB stars are needed to verify 
our synthetic model predictions. The super-AGB models of C. 
Doherty et al. (2012, in preparation) and 
H. Lau et al. (2012, in preparation) cover a larger range of mass, 
metallicities, and mass-loss rates than we have explored here. 
It will be interesting to see if these models will be able to 
produce [Rb/Fe] abundances of $\gtrsim 2$ at solar metallicity.

We can now also address the question as to whether delaying the
superwind can produce enough Rb to match the [Rb/Fe] ratios derived
for the Galactic OH/IR stars. Without the synthetic extension, the
6$\Msun$ and 7$\Msun$ models with a delayed superwind produce the
highest [Rb/Fe] ratios of 1.14 and 1.02~dex, respectively. Both of
these models are within the observed range of most Rb-rich stars
which have [Rb/Fe] $\lesssim 1.6 \pm 0.8$~dex. The synthetic
extension to the nucleosynthesis calculations results in 
[Rb/Fe] ratios of 1.3 and 1.34~dex for the 7$\Msun$ and 
6$\Msun$ models, respectively. This
means that the 6$\Msun$ with a delayed superwind can match the 
composition of the two Rb-enriched stars that have [Rb/Fe] 
$\approx 2$~dex after allowing for the substantial uncertainties 
of $\pm 0.8$~dex (only one other star has a higher ratio at
[Rb/Fe] = 2.5 dex).  Overall, we conclude that delaying the 
superwind leads to higher [Rb/Fe] ratios,
with most of the Rb-enriched stars (with [Rb/Fe] $\lesssim 1.6$~dex) 
explained by the new stellar evolution models,
even without allowing for the synthetic extension. It is also
satisfying that this simple solution is motivated by observations 
of intermediate-mass AGB stars \citep[e.g.][]{vw93,garcia07a}. 

The synthetic extension (dashed lines in Fig.~\ref{fig7}) 
that accounts for final remaining TPs after the end of HBB 
allows us to reach [Rb/Fe] abundances of $\approx 1.4$~dex. 
The increase in heavy elements comes with a substantial
increase in \iso{12}C and in all cases the final predicted C/O $> 1$.
All our model stars would be observed as dust-enshrouded
bright C-rich AGB stars \citep[as seen in the
LMC,][]{vanloon99a}. In all cases the synthetic extension takes place
when the pulsation period is very large (see Fig.~\ref{fig8}) 
and the strong variability present on the AGB has almost ceased.
Interestingly, many dust-enshrouded early post-AGB stars 
(with no signs of strong variability or very large periods)
 are found to show an inner C-rich dust shell together with an 
outer O-rich dust envelope, which is consistent with a
late change of the chemistry of the central star 
\citep[see][]{bunzel09}. These authors concluded that
because all of the post-AGB stars still host masers in an 
O-rich environment the conversion of the inner shell 
chemistry cannot have happened more than a few thousand 
years ago. All stars must have experienced the conversion 
into a C star while they were losing the last of their
envelopes before exposing their cores.

For the 6 and 7$\Msun$ models, only 4 and 7 extra TPs and TDU 
episodes are needed to reach [Rb/Fe] abundances of 1.4~dex.
These synthetic estimates are produced with a much reduced 
envelope mass owing to the effect of the luminosity-driven
superwind. The smaller dilution leads to larger [Rb/Fe] more
quickly than during the normal evolution but it also perhaps 
suggests that we are too modest with our delay in the
onset of the superwind. For some stars the superwind
may not begin until the pulsation period reaches $P \gtrsim
1000$ days or higher and this could lead to [Rb/Fe] $\gtrsim 1.4$~dex.
This explanation could also account for the composition of the
three most Rb-enriched stars, with [Rb/Fe] $\gtrsim 2$~dex.

We  studied the inclusion of a \iso{13}C-rich region 
in the He-intershell of a 5$\Msun$, $Z=0.02$ model and found it to
increase the [Rb/Fe] ratio by $\approx 0.2$~dex. This suggests 
that we may be able to obtain the average observed Rb value by 1) 
delaying the onset of the superwind even further than done here, and/or 
2) including a \iso{13}C pocket.
There are questions about the formation of \iso{13}C pockets
in intermediate-mass AGB models.  This is because 
the temperature at the base of the convective envelope during the TDU
may become hot enough for proton-capture nucleosynthesis 
\citep{goriely04,herwig04a}. For this study the most important
consequence is the likely inhibition of formation of the \iso{13}C pocket 
\citep{goriely04}. The models of \citet{herwig04a} have very deep
TDU into the CO core, as a consequence of  the diffusive convective
overshoot scheme that he employs \citep[see, e.g.,][]{herwig00}. 
One interesting consequence of this overshoot is  that each 
subsequent TP will be ignited deeper in the core and will be hotter.
This would lead to a stronger activation of the 
\iso{22}Ne($\alpha$,n)\iso{25}Mg reaction,
and more Rb production. Also, the very deep dredge-up could reach
the Rb left in the core, further enhancing the surface composition.
It remains to be seen if this can produce Rb compositions as 
high as observed in the OH/IR stars.

While it has been assumed that the metallicities of the Galactic 
OH/IR stars are solar \citep{garcia07a}, it cannot be discounted 
that there may be a small spread in metallicity.
In \citet{vanraai12} and from Fig.~\ref{fig4} we see that a reduction
in the initial metallicity results in a substantial increase in the
production of neutron-capture elements and of Rb, in particular. 
For example, using models from \citet{karakas10a} and the 172 species
network that uses the latest \citet{iliadis10} rate,  we find that a 
6$\Msun$, $Z=0.008$ ([Fe/H] =$-0.3$) model produces [Rb/Fe] =
$1.05$~dex, and a 4$\Msun$, $Z=0.004$ ([Fe/H] = $-0.7$) model makes
[Rb/Fe] = 1.02 dex. While these [Rb/Fe] abundances 
are much lower than the derived [Rb/Fe]
values of the Rb-rich stars in the Large and Small Magellanic Cloud 
\citep[up to 5~dex,][]{garcia09}, they are within the abundances 
derived for the Galactic OH/IR stars. Perhaps the OH/IR stars with
the highest [Rb/Fe] abundances of $\gtrsim 2$ evolved from slightly 
lower metallicity AGB stars of 6-7$\Msun$, [Fe/H] $\gtrsim -0.3$ 
and a delayed superwind? Observational evidence for this hypothesis 
comes from the metallicity distribution of Galactic planetary nebulae 
(PNe) that span a range from  $0.1-2.0 Z_{\odot}$, according 
to their O and Ar abundances \citep{sterling08}.  
Note that this is also the case for Type I
PNe, which are known to be the descendants of more massive AGB
stars, according to their spatial distribution and kinematics
\citep[e.g.,][]{corradi95}.
Also, a large spread of metallicities is known in other galaxies 
such as the Large Magellanic Cloud \citep[e.g.,][]{grocholski06}.
In summary, while the solutions proposed to explain the three most Rb-enriched
stars seem promising, further detailed calculations are needed
to test these scenarios.

In \citet{vanraai12} we provided a detailed discussion of the 
serious uncertainties affecting the model atmospheres used 
to derive abundances for the OH/IR AGB stars, including NLTE and
3D effects on the formation of Rb lines in the atmospheres. 
Effects caused by using 1D model atmospheres compared to 3D are
expected to be lower in the Galactic sources than in the Magellanic
Cloud AGB stars owing to their higher metallicities.  This means that 
our new results may bring observations and models into agreement 
without having to invoke large NLTE corrections. Note also
that the NLTE corrections are expected to be larger in the outliers (the
three objects with [Rb/Fe] $\gtrsim 2$~dex),
because the lines are more saturated. Finally, we do not repeat
the discussion in \citet{vanraai12} relating to the problem of the
[Rb/Zr] ratio in the model stars compared to the observations.
The problem stems from the fact that the models that produce Rb also
produce some Zr, whereas Zr is absent from the atmosphere of the 
observed AGB stars.  Here we note that the new models produce 
higher [Rb/Zr] ratios, in better agreement with the observations, 
but it still seems likely that an important fraction of the freshly 
manufactured Zr condenses into dust. 

\section{Concluding remarks} \label{sec:conclusion}

We have shown that delaying the superwind in intermediate-mass AGB
stars results in far greater production of Rb 
and other neutron-capture elements at solar metallicity. 
This indicates a possible solution to the discrepancy between
Rb observations and theoretical model predictions.
Including a synthetic
model to account for the final remaining TPs further increases the
Rb composition of the envelope. The extra Rb comes with an increase
of C and the final C/O ratio is $\ge 1$ for all models. However this
exercise shows that not many more TPs are needed for the 6$\Msun$ and
7$\Msun$ models to match the average [Rb/Fe] ratio of most of the
Rb-rich OH/IR stars.  This could be achieved by
delaying the start of the superwind mass-loss phase even further 
than done here or including a small \iso{13}C pocket. 
These solutions may also allow us to reach the [Rb/Fe]
abundances of the most Rb-rich stars with [Rb/Fe] $\gtrsim 2$.

In regards to reaction rates, we tested different nuclear networks 
and different rates for the \iso{22}Ne($\alpha$,n)\iso{25}Mg and
\iso{22}Ne($\alpha$,$\gamma$)\iso{26}Mg
reactions. Tests with a full $s$-process network of 320 species shows
that at solar metallicity the stars are not expected to produce many
elements beyond the first $s$-process peak at Rb-Sr (except perhaps
in the case when a \iso{13}C pocket is included). In regards to the
\iso{22}Ne $+ \alpha$ reactions, the NACRE rates result in the 
highest Rb abundances. The newer \iso{22}Ne $+ \alpha$ rates from
\citet{iliadis10}, which have significantly lower uncertainties, 
produce [Rb/Fe] ratios that are lower by up to 0.3~dex, whereas
the rates from \citet{karakas06a} produce final [Rb/Fe] that 
are up to 0.5~dex lower. Note that this is mostly because the 
\iso{22}Ne($\alpha$,$\gamma$)\iso{26}Mg rate from \citet{iliadis10}
has been significantly reduced, while the
\iso{22}Ne($\alpha$,n)\iso{25}Mg rate is similar to \citet{karakas06a}.
Using the \iso{22}Ne $+ \alpha$ reactions from \citet{iliadis10}
produces final maximum [Rb/Fe] $\approx 0.9$~dex and covers the
lower end of the distribution of [Rb/Fe] in the observed OH/IR AGB
stars.  To match the composition of most of the Rb-rich stars 
requires us to use the faster NACRE rates. 
This may provide an important constraint on these notoriously 
difficult to estimate reaction rates. The main issue with this
at the moment is the large uncertainties on the Rb abundances
derived from observations.

We finish with a discussion of the impact of the nucleosynthesis of
the brightest AGB stars on the chemical evolution of galaxies and
stellar systems. Delaying the superwind in intermediate-mass stars
results in a far greater production of neutron-capture elements at
solar metallicity and we would expect that this increased production
to carry on through to the lowest metallicities. Our models and the
observations by \citet{garcia06} and \citet{garcia09} show that the
most massive C-O core AGB stars of solar metallicity (up to 7$\Msun$) 
and at the metallicities of the Large and Small Magellanic Clouds 
experience considerable dredge-up. This has some important
consequences. If intermediate-mass stars at the
metallicities of the Galactic globular clusters (GCs; $-2.3
\lesssim$[Fe/H]$\lesssim -0.7$) also experience deep TDU, 
then intermediate-mass AGB stars would {\em not} be good candidates to
explain the abundance anomalies observed in every well studied GCs 
\citep[e.g.,][]{gratton04}. This has been explored quantitatively by
\citet{fenner04} and \citet{karakas06b} using yields from models with
deep dredge-up, similar to those presented here. Intermediate-mass 
stars of up to $6-7\Msun$ might instead be candidates for producing the
neutron-capture elements in some GCs, including M4
\citep{yong08a,yong08b} and M22 \citep[in particular see discussion
in][]{roederer11}.  Note that the neutron-capture elements observed 
in GCs do not show any
correlation with the light-element abundance patterns, ruling out a
relation between the two. However, models of stars with
masses near the limit of C-O core production (here the 8$\Msun$ model)
or super-AGB stars with O-Ne cores may not experience very efficient
TDU \citep[see the models of][]{siess10,ventura11}.
These objects experience very hot HBB and depending upon the model, 
 little pollution from He-intershell material (notably primary 
\iso{12}C). \citet{siess10} and \citet{ventura11} explored
this idea using super-AGB models of the appropriate metallicity. 
Further, our results and theirs show that super-AGB stars can 
produce the high helium abundances of $Y \approx 0.4$
needed to explain the multiple-populations observed in the
color-magnitude diagrams of some GCs \citep[e.g.,][]{piotto05}. 

How can we test that intermediate-mass AGB stars of low metallicity
did (or did not) produce the neutron-capture elements in GCs?
Unfortunately it is not possible to verify this production directly with 
observations today, although there do exist bright $s$-process rich 
AGB stars in the SMC with metallicities of [Fe/H] $= -0.7$ \citep{garcia09},
which are at the upper end of the GC metallicity distribution 
\citep[e.g., 47 Tucanae,][]{harris96}.
However, the most metal-poor, massive AGB stars would have evolved away a 
long time ago. One hope is that future large-scale surveys of halo stars
(e.g., the High Efficiency and Resolution Multi-Element Spectrograph 
for the AAT or SkyMapper) might reveal hints of such nucleosynthesis 
among the carbon and nitrogen-enhanced metal-poor star population. 

\acknowledgments

A.I.K. is grateful for the support of the NCI National Facility at the
ANU, and warmly thanks Arturo Manchado for hosting her visit to the IAC in
Tenerife where this work was started.  AIK also thanks the ARC for
support through a Future Fellowship (FT110100475).
D.A.G.H. acknowledges support provided by the Spanish Ministry of Science
and Innovation (MICINN) under a 2008 JdC grant and grant AYA-2007-64748.
M.L. thanks the ARC for
support through a Future Fellowship (FT100100305) and Monash
University for support through a Monash Research Fellowship.

\begin{sidewaystable}
 \begin{center}
  \caption{Characteristic parameters of the AGB models.
 \label{table1}}
  \vspace{1mm}
   \begin{tabular}{lcccccccccccc}
   \tableline\tableline
$M^{\rm i}$ & $M^{\rm f}$ & $M^{\rm f}_{\rm env}$ & $P$ & TPs & NTPs &
$\tau_{\rm ip}^{\rm f}$ & $\lambda_{\rm max}$ & $M_{\rm dredge}$ & $T_{\rm bce}^{\rm max}$ &
$T_{\rm Heshell}^{\rm max}$ & $M_{\rm bol}^{\rm
peak}$ &  $M_{\rm bol}^{\rm AGB-tip}$  \\
($\Msun$) & ($\Msun$) & ($\Msun$) & (days) & &  & (years) & & ($\Msun$) & (MK) &
(MK) &  & \\
\tableline
5.0 & 1.787 & 0.914 & 500 & 25 & 17 & 1.29($10^{4}$) & 0.95 & 6.47($10^{-2}$) & 57.3 & 344 &
$-6.32$ & $-6.21$ \\
5.0 & 1.641 & 0.756 & 700 & 72 & 64 & 1.18($10^{4}$) & 0.95 &  2.22($10^{-1}$) & 73.2 & 354 & 
$-6.51$ & $-6.29$ \\
6.0 & 1.673 & 0.761 & 500 & 33 & 25 & 7.87($10^{3}$) & 0.93 & 6.67($10^{-2}$) & 76.5 & 351 & 
$-6.58$ & $-6.34$ \\
6.0$^{\rm a}$ & 1.583 & 0.670 & 500 & 36 & 25 &  7.77($10^{3}$) & 0.95
& 6.71($10^{-2}$) & 87.2 & 355 & $-6.92$ & $-6.41$ \\
6.0 & 1.628 & 0.691 & 800 & 158 & 150 & 4.77($10^{3}$) & 0.92 & 3.56($10^{-1}$) & 84.1 & 366 &
$-$6.90 & $-$6.45 \\
7.0 & 2.150 & 1.181 & 500 & 42 & 33 & 4.67($10^{3}$) & 0.90 & 5.32($10^{-2}$) & 87.7 & 364 & 
$-$6.83 & $-$6.48 \\
7.0 & 2.115 & 1.122 & 800 & 188 & 178 & 3.50($10^{3}$) & 0.90 & 2.70($10^{-1}$) & 91.4 & 378 &
$-$7.04 & $-$6.60 \\
8.0 & 2.621 & 1.568 & 500 & 59 & 49 & 2.20($10^{3}$) & 0.89 & 3.95($10^{-2}$) & 96.8 & 378 & 
$-$7.12 & $-$6.64 \\
9.0 & 3.208 & 2.019 & 500 & 163 & 155 & 5.37($10^{2}$) & 0.77 & 3.27($10^{-2}$) & 111 & 402 & 
$-$7.55 & $-$6.89 \\
\tableline \tableline
 \end{tabular} 
Note: MK denotes the temperature in $\times 10^{6}$K,
NTPs denotes the number of TPs with peak temperatures over 300$\times 10^{6}$K,
$M_{\rm bol}^{\rm peak}$ denotes the peak AGB bolometric
luminosity, and $M_{\rm bol}^{\rm AGB-tip}$ denotes the 
AGB-tip bolometric luminosity.
(a) Computed with $\alpha = 2.5$. All other models have $\alpha = 1.86$.
 \end{center}
\end{sidewaystable}


\begin{center}
\begin{sidewaystable}
\caption{Selected key elemental abundances in [X/Fe] (in dex) and
isotopic ratios (by number) at the stellar
surface at the end of the computed evolution for the nucleosynthesis
models. Also included is the average mass fraction of helium in the
stellar wind, $<$X(He)$>$.}
\label{table2}
  \begin{tabular}{ c c c c c c c c c c c c c }
    \hline
Mass & $P$ & Net & \iso{22}Ne & $<$X(He)$>$ & C/O & \iso{12}C/\iso{13}C & [N/Fe] & 
[O/Fe] & [Na/Fe] & [Al/Fe] & \iso{24}Mg/\iso{25}Mg & \iso{24}Mg/\iso{26}Mg \\
\tableline \tableline
5.0 & 500 & 166 & K06 & 0.321 & 1.204 & 14.00 & 0.505 & $-0.037$ & 0.227 & 
0.037 & 5.524 & 4.638 \\ 
5.0 & 500 & 172 & NAC & 0.316 & 1.192 & 13.46 & 0.509 & $-0.034$ & 0.213 & 
0.057 & 5.108 & 4.213 \\
5.0 & 700 & 166 & K06 & 0.332 & 0.772 & 11.14 & 1.228 & $-0.045$ & 1.229 & 
0.114 & 2.404 & 1.700 \\
5.0 & 700 & 172 & NAC & 0.331 & 0.717 & 11.49 & 1.245 & $-0.042$ & 0.308 & 
0.157 & 2.013 & 1.378 \\
5.0 & 700 & 172 & I10 & 0.331 & 0.718 & 11.43 & 1.242 & $-0.045$ & 0.306 & 
0.140 & 2.467 & 2.828 \\
5.0 & 700 & 320 & NAC & 0.334 & 0.720 & 11.13 & 1.245 & $-0.044$ & 0.311 & 
0.160 & 1.981 & 1.375 \\ \tableline
6.0 & 500 & 166 & K06 & 0.341 & 0.678 & 14.25 & 0.869 & $-0.055$ & 0.699 & 
0.053 & 5.220 & 4.103 \\
6.0 & 500 & 172 & NAC & 0.337 & 0.647 & 14.71 & 0.879 & $-0.052$ & 0.235 & 
0.062 & 4.750 & 3.677 \\
6.0 & 500 & 320 & NAC & 0.338 & 0.633 & 13.48 & 0.884 & $-0.053$ & 0.235 & 
0.062 & 4.718 & 3.682 \\
6.0 & 800 & 166 & K06 & 0.384 & 0.887 & 8.440 & 1.410 & $-0.237$ & 1.466 & 
0.215 & 1.679 & 1.157 \\
6.0 & 800 & 172 & NAC & 0.371 & 0.711 & 9.046 & 1.423 & $-0.216$ & 0.343 & 
0.251 & 1.067 & 0.723 \\
6.0 & 800 & 172 & I10 & 0.371 & 0.710 & 8.950 & 1.420 & $-0.221$ & 0.358 & 
0.200 & 1.263 & 1.514 \\ \tableline
7.0 & 500 & 166 & K06 & 0.362 & 0.582 & 12.43 & 0.897 & $-0.100$ & 0.744 & 
0.050 & 4.734 & 3.996 \\
7.0 & 500 & 172 & NAC & 0.355 & 0.545 & 12.52 & 0.908 & $-0.096$ & 0.251 & 
0.044 & 4.144 & 3.510 \\
7.0 & 800 & 166 & K06 & 0.396 & 0.943 & 7.588 & 1.326 & $-0.375$ & 0.985 & 
0.174 & 0.714 & 0.924 \\
7.0 & 800 & 172 & NAC & 0.383 & 0.761 & 8.005 & 1.339 & $-0.361$ & 0.097 & 
0.208 & 0.360 & 0.426 \\
7.0 & 800 & 172 & I10 & 0.383 & 0.762 & 7.989 & 1.337 & $-0.363$ & 0.121 & 
0.157 & 0.386 & 0.768 \\ \tableline
8.0 & 500 & 166 & K06 & 0.367 & 0.440 & 9.462 & 0.903 & $-0.117$ & 0.623 & 
0.030 & 2.139 & 3.186 \\
8.0 & 500 & 172 & NAC & 0.366 & 0.395 & 9.282 & 0.921 & $-0.127$ & 0.208 & 
0.026 & 1.676 & 2.620 \\
8.0 & 500 & 172 & I10 & 0.362 & 0.399 & 9.506 & 0.914 & $-0.114$ & 0.188 & 
0.022 & 1.690 & 3.474 \\
8.0 & 500 & 320 & NAC & 0.361 & 0.391 & 9.342 & 0.914 & $-0.108$ & 0.171 & 
0.027 & 1.569 & 2.545 \\ \tableline
9.0 & 500 & 166 & K02 & 0.387 & 0.342 & 6.555 & 0.962 & $-0.131$ & 0.495 & 
0.02 & 0.357 & 1.114 \\
9.0 & 500 & 172 & NAC & 0.375 & 0.300 & 6.370 & 0.975 & $-0.136$ &
0.085 & 0.028 & 0.197 & 0.631 \\
\tableline \tableline
  \end{tabular}
Note: Net denotes the network used for the
calculation. The initial C/O ratio = 0.55, the initial
\iso{12}C/\iso{13}C = 89.4, and the initial \iso{24}Mg/\iso{25}Mg and
\iso{24}Mg/\iso{26}Mg ratios are 7.89 and 7.17, respectively.
Here \iso{22}Ne denotes the rate used in the calculation
where K06 is the rate from \citet{karakas06a}, NAC is the 
NACRE rate, and IL10 the rate from \citet{iliadis10}.
  \end{sidewaystable}
\end{center}

\begin{center}
\begin{sidewaystable}
\caption{Selected key heavy element abundances [X/Fe] (in dex) at the stellar surface at the 
end of the computed evolution for the nucleosynthesis models. }
\label{table3}
  \begin{tabular}{ c c c c c c c c c c c c }
    \hline
Mass & $P$ & Net & \iso{22}Ne & [Co/Fe] & [Cu/Fe] & [Se/Fe] & [Kr/Fe] & [Rb/Fe]
& [Sr/Fe] & [Zr/Fe] & [Ba/Fe] \\
\tableline \tableline
5.0 & 500 & 166 & K06 & 0.174 & 0.092 & $-0.02$ & 0.016 & 0.030 &
0.012 & $-0.077$ & -- \\
5.0 & 500 & 172 & NAC & 0.259 & 0.222 & 0.053 & 0.064 & 0.099 & 0.035
& $-0.063$ & -- \\
5.0 & 700 & 166 & K06 & 0.508 & 0.446 & 0.164 & 0.191 & 0.275 & 0.128
& $-0.016$ & -- \\
5.0 & 700 & 172 & NAC & 0.581 & 0.681 & 0.555 & 0.615 & 0.746 & 0.422
& 0.228 & -- \\
5.0 & 700 & 172 & I10 & 0.558 & 0.558 & 0.338 & 0.346 & 0.457 & 0.214
& 0.058 & -- \\
5.0 & 700 & 320 & NAC & 0.584 & 0.667 & 0.574 & 0.652 & 0.763 & 0.497
& 0.332 & 0.080 \\ \tableline
6.0 & 500 & 166 & K06 & 0.220 & 0.146 & $-0.002$ & 0.032 & 0.052 &
0.022 & $-0.072$ & -- \\
6.0 & 500 & 172 & NAC & 0.288 & 0.293 & 0.132 & 0.139 & 0.195 & 0.078
& $-0.038$ & -- \\
6.0 & 500 & 320 & NAC & 0.287 & 0.282 & 0.133 & 0.157 & 0.198 & 0.096
& 0.045 & 0.018 \\
6.0 & 800 & 166 & K06 & 0.628 & 0.671 & 0.420 & 0.492 & 0.635 & 0.387
& 0.166 & -- \\
6.0 & 800 & 172 & NAC & 0.639 & 0.870 & 0.860 & 1.026 & 1.148 & 0.807
& 0.631 & -- \\
6.0 & 800 & 172 & I10 & 0.648 & 0.775 & 0.668 & 0.764 & 0.886 & 0.563
& 0.376 & -- \\ \tableline
7.0 & 500 & 166 & K06 & 0.216 & 0.168 & 0.016 & 0.049 & 0.074 & 0.033
& $-0.068$ & -- \\
7.0 & 500 & 172 & NAC & 0.248 & 0.296 & 0.175 & 0.201 & 0.266 & 0.116
& $-0.011$ & -- \\
7.0 & 800 & 166 & K06 & 0.513 & 0.581 & 0.361 & 0.436 & 0.571 & 0.340
& 0.132 & -- \\
7.0 & 800 & 172 & NAC & 0.498 & 0.751 & 0.741 & 0.904 & 1.021 & 0.693
& 0.522 & -- \\
7.0 & 800 & 172 & I10 & 0.517 & 0.678 & 0.592 & 0.696 & 0.808 & 0.504
& 0.327 & -- \\ \tableline
8.0 & 500 & 166 & K06 & 0.197 & 0.202 & 0.054 & 0.091 & 0.132 & 0.060
& $-0.055$ & -- \\
8.0 & 500 & 172 & NAC & 0.183 & 0.300 & 0.221 & 0.284 & 0.364 & 0.172
& 0.034 & -- \\
8.0 & 500 & 172 & I10 & 0.194 & 0.268 & 0.157 & 0.194 & 0.256 & 0.112
& $-0.012$ & -- \\
8.0 & 500 & 320 & NAC & 0.195 & 0.303 & 0.242 & 0.322 & 0.397 & 0.224
& 0.133 & 0.022 \\ \tableline
9.0 & 500 & 166 & K06 & 0.155 & 0.246 & 0.106 & 0.164 & 0.242 & 0.116 
& 0.070 & -- \\
9.0 & 500 & 172 & NAC & 0.121 & 0.317 & 0.269 & 0.363 & 0.460 & 0.225
& 0.171 & -- \\
\tableline \tableline
  \end{tabular}
  \end{sidewaystable}
\end{center}

\begin{table}
 \begin{center}
  \caption{Details and results of synthetic evolution calculations.
 \label{table4}}
  \vspace{1mm}
   \begin{tabular}{lccccc}
   \tableline\tableline
Mass & $P$ & TPs remaining & $\Delta M$ & X(Rb)$_{\rm intershell}$ &
Final [Rb/Fe] \\ 
($\Msun$) & (days) &   & ($\Msun$) &   & (dex)  \\
\tableline
5.0 & 500 & 3 & 0.417 & 5.5($10^{-7}$) & 0.38 \\ 
5.0 & 700 & 2 & 0.412 & 5.6($10^{-6}$) & 0.96  \\
6.0 & 500 & 3 & 0.187 & 1.1($10^{-6}$) & 0.41 \\
6.0 & 800 & 4 & 0.193 & 7.0($10^{-6}$) & 1.34  \\ 
7.0 & 500 & 6 & 0.194 & 2.0($10^{-6}$) & 0.54 \\
7.0 & 800 & 7 & 0.150 & 8.0($10^{-6}$) & 1.26  \\
8.0 & 500 & 15 & 0.106 & 3.1($10^{-6}$) & 0.80    \\
9.0 & 500 & 60 & 0.031 & 5.3($10^{-6}$) & 0.90 \\
\tableline \tableline
  \end{tabular} 
\\
Note: Intershell compositions are from the 172 species network 
which include the NACRE rate for the \iso{22}Ne neutron source.
 \end{center}
\end{table}

\clearpage







\end{document}